\documentclass[prd,aps,a4paper,nofootinbib,twocolumn]{revtex4}  %-1

\newif\ifusesec
\usesectrue  
   
\usepackage{graphicx} 
\usepackage{mathrsfs}
\usepackage{amsmath,amsfonts,amssymb}
\usepackage{multirow}

\newcommand{\beq}{\begin{equation}}
\newcommand{\eeq}{\end{equation}}
\begin{document}

\title{New gravitational self-force analytical results for eccentric equatorial orbits around a Kerr black hole: redshift invariant}

\author{Donato \surname{Bini}}
\author{Andrea \surname{Geralico}}
\affiliation{Istituto per le Applicazioni del Calcolo ``M. Picone,'' CNR, I-00185 Rome, Italy}

\date{\today}

\begin{abstract}
The Detweiler-Barack-Sago redshift function for particles moving along slightly eccentric equatorial orbits around a Kerr black hole is currently known up to the second order in eccentricity, second order in spin parameter, and the 8.5 post-Newtonian order.
We improve the analytical computation of such a gauge-invariant quantity by including  terms up to the fourth order in eccentricity at the same post-Newtonian approximation level.
We also check that our results agrees with the corresponding post-Newtonian expectation of the same quantity, calculated by using the currently known Hamiltonian for spinning binaries.
\end{abstract}

%\pacs{04.20.Cv}
%\keywords{eccentric orbits, Kerr black holes, redshift invariant}
\maketitle

\section{Introduction}

Low-frequency gravitational wave signals from binary systems with a very small mass ratio are expected to be detected by planned space-based gravitational wave observatories, such as the forthcoming eLISA \cite{elisa}. The dynamics of such systems is well described by black hole perturbation theory within the gravitational self-force (GSF) approach. According to this formalism, the motion of the smaller body can be treated as a perturbation of the background gravitational field of the larger body to the linear order in their mass ratio. GSF calculations require advanced mathematical tools to reconstruct the metric perturbation, whose components diverge at the particle's location, so that a suitable regularization procedure is needed to isolate their finite contribution (see, e.g., Ref \cite{Barack:2009ux}).

The major contribution of GSF in the last few years has been the computation of several gauge-invariant quantities, which can be used to compare results between different approximation methods in the overlapping regime of validity. Furthermore, this allows one to validate and inform Post-Newtonian (PN) techniques and numerical relativity (NR) simulations as well as to calibrate the Effective-One-Body (EOB) model \cite{Buonanno:1998gg,Buonanno:2000ef,Damour:2001tu}.
The first such invariant to be calculated was the linear-in-mass-ratio change in the coordinate time component of the particle's 4-velocity, or redshift invariant, on a circular orbit around a Schwarzschild black hole, introduced by Detweiler \cite{Detweiler:2008ft}.
A complete methodology to perform analytic high PN order self-force computations was developed by Bini and Damour \cite{Bini:2013zaa} in the framework of Regge-Wheeler-Zerilli \cite{Regge:1957td,Zerilli:1971wd} (RWZ) formalism, allowing them to calculate the redshift invariant at the 9.5 PN level \cite{Bini:2015bla}, soon after pushed at the 22.5PN one by Kavanagh and collaborators \cite{Kavanagh:2015lva}. 
The inclusion of the rotation of the background spacetime is first due to Shah, who computed the redshift invariant along circular orbits in Kerr spacetime at 4PN order \cite{shah_capra2015,shah_MG14} by using the Teukolsky formalism and a radiation gauge \cite{Shah:2012gu}, further improved in Refs. \cite{Bini:2015xua,Kavanagh:2016idg}.

The generalization to slightly eccentric orbits was discussed by Barack and Sago \cite{Barack:2011ed} still in the case of a non-rotating black hole, who calculated the orbit-averaged value of the redshift invariant for given azimuthal and radial frequencies by using a Lorenz gauge, hereafter the Detweiler-Barack-Sago (DBS) redshift function. 
High-PN calculations were done in Refs. \cite{Bini:2015bfb,Bini:2016qtx} up to the fourth order in the eccentricity. Higher order terms in the eccentricity were obtained in Refs. \cite{Hopper:2015icj,Bini:2016qtx}, but at the 4PN level of approximation only.
The first analytic computation of the self-force correction to the DBS redshift function for a small mass in eccentric equatorial orbit around a Kerr black hole was done in Ref. \cite{Bini:2016dvs}, following the standard Teukolsky perturbation scheme. The results presented there gave the redshift contributions mixing eccentricity and spin effects through second order in both eccentricity and spin parameter, and were accurate to the 8.5 PN order. 
Here we improve this computation by including terms which are fourth order in the eccentricity at the same PN approximation level.
We also calculate the corresponding comparable-mass redshift by using the current knowledge of the Arnowitt-Deser-Misner (ADM) Hamiltonian for two point masses with aligned spins \cite{Schafer:2018kuf}, providing an independent check of the first few PN orders in our results. 

The masses of the two bodies are denoted by $m_1$ and $m_2$, with the convention that $m_1\le m_2$.
We define, in a standard way, the mass ratio $q={m_1}/{m_2}$, the reduced mass $\mu ={m_1m_2}/{M}$ and the symmetric mass ratio $\nu ={\mu}/{M}$, with $M=m_1+m_2$ the total mass, and the reduced mass difference $\Delta=(m_2-m_1)/M=\sqrt{1-4\nu}$. 
The bodies are endowed with spin, denoted by $S_1$ and $S_2$, respectively.
We also introduce the dimensionless spin variables $\chi_{1,2}\equiv S_{1,2}/m_{1,2}^2$ associated with each body.
GSF results are obtained in the limit of small mass-ratio ($q\sim\nu\ll 1$) and vanishing spin $S_1=0$ of the smaller body.
We closely follow the notation and convention of Ref. \cite{Bini:2016dvs}. 
The metric signature is chosen to be $-2$ and units are such that $c=G=1$ unless differently specified.
Greek indices run from 0 to 3, whereas Latin ones from 1 to 3.

\section{Perturbations on a Kerr spacetime}

The background Kerr metric with parameters $m_2$ and $a_2=a$ (with $\hat a =a/m_2$ dimensionless) written in Boyer-Lindquist coordinates $(t,r,\theta,\phi)$ reads
\begin{eqnarray}
\label{kerrmet}
ds_{(0)}^2&=&g^{(0)}_{\alpha\beta}dx^\alpha dx^\beta\nonumber\\
&=&\left(1-\frac{2m_2r}{\Sigma}  \right) dt^2+\frac{4am_2r \sin^2\theta}{\Sigma}dtd\phi\nonumber\\
&-& \frac{\Sigma}{\Delta}dr^2-\Sigma d\theta^2\nonumber\\
&-&  \left( r^2+a^2+\frac{2m_2ra^2\sin^2\theta}{\Sigma} \right)\sin^2\theta d\phi^2\,,
\end{eqnarray}
where 
\beq
\Delta= r^2+a^2-2m_2r\,,\qquad 
\Sigma=r^2+a^2\cos^2\theta\,.
\eeq
Let the perturbation be associated with a particle of mass $m_1$ moving along a slightly eccentric equatorial geodesic orbit, with four velocity $u^\mu\equiv u^t k^\mu$, $u^t=dt/d\tau$ and $k^\mu\equiv \partial_t +(dr/dt)\partial_r +(d\phi/dt) \partial_\phi$.
It is convenient to parametrize the orbit in terms of eccentricity $0\le e<1$ and semi-latus rectum $m_2p$ so that  
\beq\label{rversuschi}
r =\frac{m_2p}{1+e\cos \chi}\,,
\eeq
where $\chi\in[0,2\pi]$, with $p$ (as well as its reciprocal $u_p=1/p$) dimensionless. 
The orbit thus oscillates between a minimum radius $r_{\rm peri}$ ($\chi=0$, periastron) and a maximum radius $r_{\rm apo}$ ($\chi=\pi$, apastron).
The background motion is governed by the following equations \cite{Glampedakis:2002ya,Bini:2016iym}

\begin{widetext}

\begin{eqnarray}
m_2 \frac{d\chi}{d\tau} &=&  u_p^{3/2}(1+e\cos \chi )^2
[1+u_p^2\, \hat x{}^2 ( e^2-2 e\cos\chi-3)]^{1/2}
\,,\nonumber\\
\frac{dt}{d\chi}&=&\frac{m_2}{u_p^{3/2}}\frac{E +E\hat a^2 u_p^2 (1+e\cos\chi)^2   -2 \hat a u_p^3\hat x (1+e\cos\chi)^3}{(1+e\cos \chi )^2[1+u_p^2 \,\hat x{}^2 (e^2-2 e\cos \chi -3) ]^{1/2}
[1-2 u_p(1+ e\cos \chi) +a^2 u_p^2(1+ e\cos \chi)^2  ]}
\,,\nonumber\\
\frac{d\phi}{d\chi}&=& u_p^{1/2}\frac{ \hat x + \hat a E - 2 u_p \hat x (1+  e\cos \chi) }{[1+u_p^2 \,\hat x{}^2 (e^2-2 e\cos \chi -3) ]^{1/2}
[1-2 u_p(1+ e\cos \chi) +a^2 u_p^2(1+ e\cos \chi)^2  ]}
\,.
\end{eqnarray}
Here $E$ and $L$ are the conserved energy and angular momentum per unit mass of the particle, so that $E$ and $L/m_2$ are dimensionless, together with their combination $\hat x=(L-aE)/m_2$. 
Their explicit expressions in terms of ($u_p, e, \hat a$) for prograde orbits are given by 
\begin{eqnarray}
E&=&\frac{1-2u_p+{\hat a}u_p^{3/2}}{\sqrt{1-3u_p+2{\hat a}u_p^{3/2}}}\bigg\{1
+u_p\frac{(1-4u_p)^2-u_p^{3/2}(-7+26u_p){\hat a}+u_p^2(-1+10u_p){\hat a}^2+3{\hat a}^3u_p^{7/2}-2{\hat a}^4u_p^4}
{2(1-3u_p+2{\hat a}u_p^{3/2})(1-2u_p+{\hat a}^2u_p^2)(1-2u_p+{\hat a}u_p^{3/2})}\,e^2\nonumber\\
&&
+\frac{u_p^2}{8(1-3u_p+2{\hat a}u_p^{3/2})^2(1-2u_p+{\hat a}^2u_p^2)^3(1-2u_p+{\hat a}u_p^{3/2})}\left[
(3-8u_p)(1-4u_p)^2-(8-111u_p+514u_p^2\right.\nonumber\\
&&
-900u_p^3+424u_p^4)u_p^{1/2}{\hat a}+(4-91u_p+626u_p^2-1620u_p^3+1304u_p^4)u_p{\hat a}^2-(-4+163u_p-924u_p^2\nonumber\\
&&
+1308u_p^3)u_p^{5/2}{\hat a}^3+(12-107u_p+180u_p^2+124u_p^3)u_p^3{\hat a}^4+(28-323u_p+670u_p^2)u_p^{9/2}{\hat a}^5\nonumber\\
&&\left.
-(-12-71u_p+478u_p^2)u_p^5{\hat a}^6+(-4+111u_p)u_p^{13/2}{\hat a}^7+4(1+u_p)u_p^7{\hat a}^8-4{\hat a}^9u_p^{17/2}
\right]\,e^4
+O(e^6)\bigg\}
\,,\nonumber\\
\end{eqnarray}
\begin{eqnarray}
\frac{L}{m_2}&=&\frac{1-2{\hat a}u_p^{3/2}+{\hat a}^2u_p^2}{\sqrt{u_p(1-3u_p+2{\hat a}u_p^{3/2})}}\bigg\{1
+\frac{u_p}
{2(1-3u_p+2{\hat a}u_p^{3/2})(1-2u_p+{\hat a}^2u_p^2)(1-2u_p+{\hat a}u_p^{3/2})}\left[
1-2u_p\right.\nonumber\\
&&\left.
+2(-1+u_p+8u_p^2)u_p^{1/2}{\hat a}-2(-1+4u_p+13u_p^2)u_p{\hat a}^2+2(4+5u_p)u_p^{5/2}{\hat a}^3+(-2+3u_p)u_p^3{\hat a}^4-2{\hat a}^5u_p^{9/2}
\right]\,e^2\nonumber\\
&&
+\frac{u_p^2}{8(1-3u_p+2{\hat a}u_p^{3/2})^2(1-2u_p+{\hat a}^2u_p^2)^3(1-2u_p+{\hat a}u_p^{3/2})}\left[
3(1-2u_p)^3-2(6-41u_p+74u_p^2-20u_p^3)u_p^{1/2}{\hat a}\right.\nonumber\\
&&
-2(-8+72u_p-109u_p^2-172u_p^3+212u_p^4)u_p{\hat a}^2+2(-4+68u_p-149u_p^2-348u_p^3+652u_p^4)u_p^{3/2}{\hat a}^3\nonumber\\
&&
-2(22-139u_p-117u_p^2+654u_p^3)u_p^3{\hat a}^4+2(-12-4u_p+83u_p^2+62u_p^3)u_p^{7/2}{\hat a}^5+2(8-70u_p+44u_p^2\nonumber\\
&&\left.
+335u_p^3)u_p^4{\hat a}^6-2(-28+116u_p+239u_p^2)u_p^{11/2}{\hat a}^7+(92+111u_p)u_p^7{\hat a}^8+4(-2+u_p)u_p^{15/2}{\hat a}^9-4{\hat a}^{10}u_p^9
\right]\,e^4\nonumber\\
&&
+O(e^6)\bigg\}
\,,
\end{eqnarray}

\end{widetext}
respectively, to the fourth order in eccentricity.

The radial and azimuthal periods and associated frequencies are
\beq
T_{r0} =\oint dt
= \int_0^{2\pi} \frac{dt}{d\chi} \, d\chi\,,\qquad
\Omega_r = \frac{2\pi}{T_{r0}}
\,,
\eeq
and
\beq
\Phi_0=\oint d\phi
= \int_0^{2\pi} \frac{d\phi}{d\chi} \, d\chi\,,\qquad
\Omega_\phi =\frac{\Phi_0}{T_{r0}}\,,
\eeq
respectively, and can be expressed in terms of elliptic integrals.
The first terms of their small-eccentricity expansion read

\begin{widetext}

\begin{eqnarray} 
\frac{T_{r0}}{2\pi m_2}&=&
\frac{1+{\hat a}u_p^{3/2}}{u_p^{3/2}\sqrt{1-6u_p+8{\hat a}u_p^{3/2}-3{\hat a}^2u_p^2}}\bigg\{
1+\frac34\frac1{(1+{\hat a}u_p^{3/2})(1-2u_p+{\hat a}^2u_p^2)(1-6u_p+8{\hat a}u_p^{3/2}-3{\hat a}^2u_p^2)^2}\nonumber\\
&&
\times 
\left[2-32u_p+165u_p^2-266u_p^3-u_p^{3/2}(-38+376u_p-841u_p^2+2u_p^3){\hat a}-u_p^2(12-314u_p+999u_p^2+16u_p^3){\hat a}^2\right.\nonumber\\
&&
+u_p^{7/2}(-108+466u_p+93u_p^2){\hat a}^3-u_p^4(-11-32u_p+176u_p^2){\hat a}^4+u_p^{11/2}(-101+160u_p){\hat a}^5\nonumber\\
&&\left.
-u_p^6(-25+72u_p){\hat a}^6+13u_p^{15/2}{\hat a}^7\right]\,e^2
+O(e^4)\bigg\}
\,,\nonumber\\
\frac{\Phi_0}{2\pi}&=&
\frac{1}{\sqrt{1-6u_p+8{\hat a}u_p^{3/2}-3{\hat a}^2u_p^2}}\bigg\{
1+\frac34u_p^2(1-{\hat a}u_p^{1/2})^3\frac{1-2u_p-(-3+22u_p)u_p^{1/2}\hat a+33{\hat a}^2u_p^2-13{\hat a}^3u_p^{5/2}}
{(1-2u_p+{\hat a}^2u_p^2)(1-6u_p+8{\hat a}u_p^{3/2}-3{\hat a}^2u_p^2)^2}\,e^2\nonumber\\
&&
+O(e^4)\bigg\}\,,
\end{eqnarray}
respectively.
Similarly, the proper time period is defined by 
\beq
{\mathcal T}_{r0}=\oint d\tau
= \int_0^{2\pi} \frac{d\tau}{d\chi} \, d\chi\,,
\eeq
with
\begin{eqnarray} 
\frac{{\mathcal T}_{r0}}{2\pi m_2}&=&
\frac{\sqrt{1-3u_p+2{\hat a}u_p^{3/2}}}{u_p^{3/2}\sqrt{1-6u_p+8{\hat a}u_p^{3/2}-3{\hat a}^2u_p^2}}\bigg\{
1+\frac34\frac1{(1-3u_p+2{\hat a}u_p^{3/2})(1-2u_p+{\hat a}^2u_p^2)(1-6u_p+8{\hat a}u_p^{3/2}-3{\hat a}^2u_p^2)^2}\nonumber\\
&&
\times 
\left[(-1+2u_p)(-2+32u_p-165u_p^2+267u_p^3)-2(-20+242u_p-879u_p^2+966u_p^3)u_p^{3/2}{\hat a}\right.\nonumber\\
&&
+(-12+376u_p-1997u_p^2+2729u_p^3)u_p^2{\hat a}^2-2(60-494u_p+837u_p^2)u_p^{7/2}{\hat a}^3+(11-95u_p+88u_p^2)u_p^4{\hat a}^4\nonumber\\
&&\left.
+2(-45+226u_p)u_p^{11/2}{\hat a}^5-(-25+233u_p)u_p^6{\hat a}^6+38{\hat a}^7u_p^{15/2}\right]\,e^2
+O(e^4)\bigg\}\,.
\end{eqnarray}

\end{widetext}
The ratio between the coordinate time period and the proper time period then defines the (unperturbed) redshift variable $U_0=T_{r0}/{\mathcal T}_{r0}$.

\subsection{Detweiler-Barack-Sago redshift function}

The DBS (inverse) redshift function $U$ is defined as \cite{Barack:2011ed}
\beq
\label{U}
U\left(m_2\Omega_r, m_2\Omega_\phi, a_2, q\right)= \frac{\displaystyle\oint dt}{\displaystyle\oint d\tau}=\frac{T_r}{{\mathcal T}_r} \, ,
\eeq
where the coordinate time and proper time radial periods now include all conservative self-force corrections referring to the perturbed spacetime metric
\begin{eqnarray}
\label{gperturbed}
g_{\mu\nu}(x^\alpha; m_1, m_2, a_2)&=&g^{(0)}_{\mu\nu}(x^\alpha; m_2, a_2)\nonumber\\
&+&
q h_{\mu\nu}(x^\alpha)+O\left(q^2\right)\,,
\end{eqnarray}
with $g^{(0)}_{\mu\nu}(x^\alpha; m_2, a_2)$ being the background metric \eqref{kerrmet} and $q h_{\mu\nu}(x^\alpha)$ the perturbation.
The (first-order) self-force contribution $\delta U$ to the function \eqref{U} is then given by the expansion
\begin{eqnarray} 
\label{UdU}
U\left(m_2\Omega_r, m_2\Omega_\phi, a_2, q\right)&=& U_0\left(m_2\Omega_r, m_2\Omega_\phi, a_2\right) \nonumber \\ 
&+&
q\delta U\left(m_2\Omega_r, m_2\Omega_\phi, a_2\right)\nonumber\\
&+&
 O(q^2)\,,
\end{eqnarray}
which is performed at fixed orbital frequencies, and it is defined in terms of the $O(q)$ metric perturbation $h_{\mu\nu}$ by the following coordinate time
average \cite{Bini:2016dvs}
\beq
\label{delta_U1}
\delta U (u_p,e, \hat a)=\frac12 \, (U_{0})^2\langle h_{uk}\rangle_{t}\,,
\eeq
where $h_{uk}=h_{\mu\nu}u^\mu k^\nu$ (equivalent to the original definition of Ref. \cite{Barack:2011ed} in terms of the proper time average of $h_{uu}=h_{\mu\nu}u^\mu u^\nu$, being $U_0\langle h_{uk}\rangle_{t}=\langle h_{uu}\rangle_{\tau}$).
Finally, it can be conveniently reexpressed in terms of the eccentricity $e$ and dimensionless (inverse) semi-latus rectum $u_p$ of the orbit.
The expansion of $\delta U(u_p, e, \hat a)$ in powers of $e$ and $\hat a$ then reads
\begin{eqnarray} 
\label{dUea}
\delta U(u_p,e,\hat a) &=&\sum_{i,j=0}^\infty e^i  {\hat a}^j \delta U^{(e^i, a^j)}(u_p) \nonumber\\
&=& \delta U^{(e^0, a^0)}+e^2 \delta U^{(e^2, a^0)}+e^4 \delta U^{(e^4, a^0)}\nonumber\\
&+& \hat a \delta U^{(e^0, a^1)}+\hat a^2\delta U^{(e^0, a^2)}+\hat a^3\delta U^{(e^0, a^3)}\nonumber\\
&+& \hat a^4\delta U^{(e^0, a^4)}+\hat a^5\delta U^{(e^0, a^5)}+\hat a^6\delta U^{(e^0, a^6)}\nonumber\\
&+& e^2 \hat a\delta U^{(e^2, a^1)}+e^2 \hat a^2\delta U^{(e^2, a^2)}\nonumber\\
&+& e^4 \hat a\delta U^{(e^4, a^1)}+e^4 \hat a^2\delta U^{(e^4, a^2)}
+\ldots\,.
\end{eqnarray}
The spin-independent part is known up to $e^{20}$, but at 4PN order only \cite{Hopper:2015icj,Bini:2016qtx}.
Higher-PN order computations were done in Refs. \cite{Bini:2015bfb,Bini:2016qtx} up to $e^4$.
The spin-dependent part mixing spin and eccentricity was computed in Ref. \cite{Bini:2016dvs} to the second order in both parameters through the 8.5PN order.
In this work we improve such a result by including the terms $\delta U^{(e^4,a^1)}(u_p)$ and $\delta U^{(e^4,a^2)}(u_p)$ which are fourth order in the eccentricity, at the same PN level.

\section{Self-force results}

For the present computation we closely follow the standard Teukolsky perturbation scheme as discussed in detail in Refs. \cite{Shah:2012gu,vandeMeent:2015lxa} and already adopted in our previous work \cite{Bini:2016dvs} (see also the Appendix A there), so we limit below to provide the necessary information on intermediate steps.
Our computed quantity $\langle h_{uk}\rangle_t$ is regularized by subtracting its PN-analytically computed large-$l$ limit (we refer, e.g., to Section IIIB of Ref. \cite{Bini:2018ylh} for a discussion on the regularization procedure of gauge-invariant quantities and related issues). We give below the subtraction term $B$ of the quantity $U_0\langle h_{uk}\rangle_t$, whose expansion is given by
\begin{eqnarray}
B(u_p, e, \hat a) &=&\sum_{i,j=0}^\infty e^i {\hat a}^j B^{(e^i,a^j)}(u_p)\nonumber\\
&=& B^{(e^0,a^0)}+e^2 B^{(e^2,a^0)}+e^4 B^{(e^4,a^0)}\nonumber\\
&+& \hat a B^{(e^0,a^1)}+\hat a^2B^{(e^0,a^2)}\nonumber\\
&+& e^2 \hat aB^{(e^2,a^1)}+e^2 \hat a^2B^{(e^2,a^2)}\nonumber\\
&+& e^4 \hat aB^{(e^4,a^1)}+e^4 \hat a^2B^{(e^4,a^2)}+\ldots\,.
\end{eqnarray}
The new coefficients relevant here are the following
\begin{widetext}
\begin{eqnarray}
-B^{(e^4,a^1)}&=& 
-\frac{5}{2}u_p^{5/2}-\frac{171}{8}u_p^{7/2}-\frac{20353}{128}u_p^{9/2}-\frac{280531}{256}u_p^{11/2}-\frac{226368825}{32768}u_p^{13/2}-\frac{5144048057}{131072}u_p^{15/2}\nonumber\\
&&
-\frac{402039445253}{2097152}u_p^{17/2}-\frac{1416360159939}{2097152}u_p^{19/2} 
+O(u_p^{21/2})
\,,\nonumber\\
-B^{(e^4,a^2)}&=&
\frac{23}{16}u_p^3+\frac{225}{8}u_p^4+\frac{172279}{512}u_p^5+\frac{12893677}{4096}u_p^6+\frac{6467356313}{262144}u_p^7+\frac{87273644687}{524288}u_p^8\nonumber\\
&&
+\frac{122581805463}{131072}u_p^9
+O(u_p^{10})
\,.
\end{eqnarray}

The non-radiative multipoles ($l=0,1$) have been computed separately, as in Eq. (138) of Ref. \cite{vandeMeent:2015lxa}. The corresponding (already subtracted) contributions to $\delta U$ are the following
\begin{eqnarray}
-\delta U^{(e^4,a^1)}_{l=0,1}&=&
\frac{7}{8}u_p^{5/2}+\frac{243}{8}u_p^{7/2}+\frac{43121}{128}u_p^{9/2}+\frac{624833}{256}u_p^{11/2}+\frac{487725977}{32768}u_p^{13/2}+\frac{10627100367}{131072}u_p^{15/2}\nonumber\\
&&
+\frac{819109245181}{2097152}u_p^{17/2}+\frac{3144664846399}{2097152}u_p^{19/2}
+O(u_p^{21/2})
\,,\nonumber\\
-\delta U^{(e^4,a^2)}_{l=0,1}&=&
-\frac{7}{16}u_p^3-\frac{1281}{32}u_p^4-\frac{261339}{512}u_p^5-\frac{20240129}{4096}u_p^6-\frac{10393992633}{262144}u_p^7-\frac{71644898835}{262144}u_p^8\nonumber\\
&&
-\frac{3350649968345}{2097152}u_p^9
+O(u_p^{10})
\,.
\end{eqnarray}

We list below the new contributions to the eccentricity-spin decomposition \eqref{dUea} of $\delta U(u_p, e, \hat a)$:
\begin{eqnarray}
-\delta U^{(e^4,a^1)}&=&C_{2.5}^{(e^4,a^1),{\rm c}}  u_p^{5/2}+C_{3.5}^{(e^4,a^1),{\rm c}}  u_p^{7/2}+C_{4.5}^{(e^4,a^1),{\rm c}}u_p^{9/2}+C_{5.5}^{(e^4,a^1),{\rm c}}u_p^{11/2}
 +\left(C_{6.5}^{(e^4,a^1),{\rm c}}+ C_{6.5}^{{(e^4,a^1)},\ln{}} \ln(u_p) \right) u_p^{13/2}\nonumber\\
&&
+\left( C_{7.5}^{(e^4,a^1),{\rm c}}+ C_{7.5}^{{(e^4,a^1)},\ln{}}\ln(u_p) \right) u_p^{15/2}
+C_{8}^{(e^4,a^1),{\rm c}}  u_p^8+\left(C_{8.5}^{(e^4,a^1),{\rm c}}+ C_{8.5}^{{(e^4,a^1)},\ln{}} \ln(u_p)\right) u_p^{17/2}\nonumber\\
&&
+C_{9}^{(e^4,a^1),{\rm c}}u_p^9
+\left(C_{9.5}^{(e^4,a^1),{\rm c}}+ C_{9.5}^{{(e^4,a^1)},\ln{}} \ln(u_p)+ C_{9.5}^{{(e^4,a^1)},\ln^2{}}\ln(u_p)^2 \right) u_p^{19/2}
+O_{\rm ln}(u_p^{10})
\,,
\end{eqnarray}
with
\begin{eqnarray}
C_{2.5}^{(e^4,a^1),{\rm c}}&=& \frac18
\,,\qquad 
C_{3.5}^{(e^4,a^1),{\rm c}}= \frac{117}{4}
\,,\qquad
C_{4.5}^{(e^4,a^1),{\rm c}}= \frac{6277}{16}
\,,\qquad
C_{5.5}^{(e^4,a^1),{\rm c}}= 3547-\frac{2025}{128}\pi^2
\,,\nonumber\\
C_{6.5}^{(e^4,a^1),{\rm c}}&=& \frac{11079823}{400}-\frac{482037}{1024}\pi^2+\frac{2496}{5}\gamma-\frac{782912}{15}\ln(2)+\frac{328779}{10}\ln(3)
\,,\qquad
C_{6.5}^{{(e^4,a^1)},\ln{}}= \frac{1248}{5}
\,,\nonumber\\
C_{7.5}^{(e^4,a^1),{\rm c}} &=& \frac{15928768049}{58800}-\frac{342977251}{24576}\pi^2+\frac{112601}{105}\gamma+\frac{19639201}{105}\ln(2)+\frac{5568831}{140}\ln(3)-\frac{5859375}{56}\ln(5)
\,,\nonumber\\
C_{7.5}^{{(e^4,a^1)},\ln{}}&=& \frac{119657}{210}
\,,\qquad
C_{8}^{(e^4,a^1),{\rm c}} = -\frac{777593}{1575}\pi
\,,\nonumber\\
C_{8.5}^{(e^4,a^1),{\rm c}} &=& -\frac{4111458343}{30240}+\frac{484156469287}{7077888}\pi^2+\frac{56533786}{2835}\gamma-\frac{6602633558}{2835}\ln(2)+\frac{4721733}{40}\ln(3)\nonumber\\
&&
 +\frac{4446484375}{4536}\ln(5)+\frac{1285031305}{1048576}\pi^4
\,,\qquad
C_{8.5}^{{(e^4,a^1)},\ln{}}= \frac{28294109}{2835}
\,,\qquad
C_{9}^{(e^4,a^1),{\rm c}} = \frac{41378241209}{4233600}\pi
\,,\nonumber\\
C_{9.5}^{(e^4,a^1),{\rm c}}&=& -\frac{1044120684387514309}{67234860000}+\frac{3272645023268729}{1651507200}\pi^2-\frac{1998265295479}{3638250}\gamma-\frac{6696061961359}{404250}\ln(2) \nonumber\\
&& 
+\frac{1342084137130899}{68992000}\ln(3)-\frac{3500960078125}{709632}\ln(5)-\frac{96889010407}{46080}\ln(7)-\frac{318096}{5}\zeta(3)\nonumber\\
&&
+\frac{4481923074363}{671088640}\pi^4+\frac{8587512}{175}\gamma^2+\frac{5072685424}{1575}\gamma\ln(2)-\frac{292450014}{175}\gamma\ln(3)+\frac{638521288}{105}\ln(2)^2\nonumber\\
&&
-\frac{292450014}{175}\ln(2)\ln(3)-\frac{146225007}{175}\ln(3)^2
\,,\nonumber\\
C_{9.5}^{{(e^4,a^1)},\ln{}}&=&-\frac{2026810797079}{7276500}+\frac{8587512}{175}\gamma+\frac{2536342712}{1575}\ln(2)-\frac{146225007}{175}\ln(3)
\,,\qquad
C_{9.5}^{{(e^4,a^1)},\ln^2{}}=\frac{2146878}{175}
\,,\nonumber\\
\end{eqnarray}
and
\begin{eqnarray}
-\delta U^{(e^4,a^2)}&=&C_{3}^{(e^4,a^2),{\rm c}}  u_p^{3}+C_{4}^{(e^4,a^2),{\rm c}}  u_p^{4}+C_{5}^{(e^4,a^2),{\rm c}} u_p^{5}
+C_{6}^{(e^4,a^2),{\rm c}} u_p^{6}
 +\left(C_{7}^{(e^4,a^2),{\rm c}}+ C_{7}^{{(e^4,a^2)},\ln{}} \ln(u_p) \right) u_p^{7}\nonumber\\
&& 
+\left( C_{8}^{(e^4,a^2),{\rm c}}+ C_{8}^{{(e^4,a^2)},\ln{}}\ln(u_p)\right) u_p^{8}
+C_{8.5}^{(e^4,a^2),{\rm c}}  u_p^{17/2}\nonumber\\
&&+\left(C_{9}^{(e^4,a^2),{\rm c}}+ C_{9}^{{(e^4,a^2)},\ln{}} \ln(u_p)\right) u_p^{9}
+C_{9.5}^{(e^4,a^2),{\rm c}} u_p^{19/2}
+O_{\rm ln}(u_p^{10})
\,,
\end{eqnarray}
with
\begin{eqnarray}
C_{3}^{(e^4,a^2),{\rm c}}&=& -\frac14
\,,\qquad  
C_{4}^{(e^4,a^2),{\rm c}}= -\frac{387}{8}
\,,\qquad
C_{5}^{(e^4,a^2),{\rm c}}= -\frac{5401}{8}
\,,\qquad
C_{6}^{(e^4,a^2),{\rm c}}= -\frac{354349}{48}+\frac{15455}{2048}\pi^2
\,,\nonumber\\
C_{7}^{(e^4,a^2),{\rm c}}&=& -\frac{4825888}{75}+\frac{291597}{4096}\pi^2-118\gamma+\frac{209846}{15}\ln(2)-\frac{349191}{40}\ln(3)
\,,\qquad
C_{7}^{{(e^4,a^2)},\ln{}}=-59
\,,\nonumber\\
C_{8}^{(e^4,a^2),{\rm c}}&=&-\frac{1557821011}{4200}-\frac{8831813359}{786432}\pi^2+\frac{2731}{15}\gamma+\frac{47764909}{105}\ln(2)-\frac{73024659}{224}\ln(3)
+\frac{21484375}{672}\ln(5)
\,,\nonumber\\
C_{8}^{{(e^4,a^2)},\ln{}}&=& \frac{2731}{30}
\,,\qquad
C_{8.5}^{(e^4,a^2),{\rm c}}= \frac{360697}{1800}\pi
\,,\nonumber\\
C_{9}^{(e^4,a^2),{\rm c}}&=& -\frac{392978504729}{88200}+\frac{1343912434839299}{9909043200}\pi^2-\frac{10714526}{2835}\gamma+\frac{953618588}{567}\ln(2)-\frac{238806549}{112}\ln(3)\nonumber\\
&& 
+\frac{763671875}{1008}\ln(5)+\frac{6264}{5}\zeta(3)-\frac{4452007537}{134217728}\pi^4
\,,\nonumber\\
C_{9}^{{(e^4,a^2)}\ln{}}&=&-\frac{1407541}{2835}
\,,\qquad
C_{9.5}^{(e^4,a^2),{\rm c}}=\frac{4291787179}{235200}\pi
\,.
\end{eqnarray}

\end{widetext}

\section{PN check}

Let us check the first PN terms of our results by using the Hamiltonian description of a two-body system with spin.
We use the center-of-mass ADM Hamiltonian, including both linear and quadratic-in-spin terms up to the present knowledge, namely next-to-next-to-leading-order (NNLO) for the linear-in-spin terms and next-to-leading-order (NLO) for the quadratic-in-spin terms (see Ref. \cite{Schafer:2018kuf} for a recent review).
We will limit ourselves to the case of two point masses with aligned spins, orthogonal to the orbital motion.

\subsection{ADM Hamiltonian}

The ADM Hamiltonian of the system reads 
\beq
\label{Hadmdef}
H^{\rm ADM}=m_1+m_2+\mu \hat H^{\rm ADM}\,,
\eeq
with
\beq
\hat H^{\rm ADM}=\hat H_{\rm orb}^{\rm ADM}+\hat H_{\rm SO}^{\rm ADM}+\hat H_{\rm SS}^{\rm ADM}
\,.
\eeq
The reduced center-of-mass Hamiltonian $\hat H^{\rm ADM}=\hat H^{\rm ADM}(R,P_r,P_\phi; m_1, m_2; S_1,S_2)$ is a function of the reduced variables $R$, $P_r$, $P_\phi$ and the masses and spins of the two bodies.
The orbital Hamiltonian $\hat H_{\rm orb}^{\rm ADM}$ is explicitly known at the 4PN level \cite{Damour:2014jta}, but for our purposes it is enough to use it through the 3PN, 
\beq
\hat H_{\rm orb}^{\rm ADM}=\hat H_{\rm orb}^{\rm  N }+\hat H_{\rm orb}^{\rm ADM, 1PN }+\hat H_{\rm orb}^{\rm ADM, 2PN }+\hat H_{\rm orb}^{\rm ADM, 3PN }\,.
\eeq
The spin-orbit (SO) Hamiltonian $\hat H_{\rm SO}^{\rm ADM}$ is explicitly known up the NNLO level,
\beq
\hat H_{\rm SO}^{\rm ADM}=\hat H_{\rm SO}^{\rm ADM,LO}+\hat H_{\rm SO}^{\rm ADM,NLO}+\hat H_{\rm SO}^{\rm ADM,NNLO}\,,
\eeq
whereas the spin-spin (SS) Hamiltonian $\hat H_{\rm SS}^{\rm ADM}$ is explicitly known up the NLO level,
\beq
\hat H_{\rm SS}^{\rm ADM}=\hat H_{\rm SS}^{\rm ADM,LO}+\hat H_{\rm SS}^{\rm ADM,NLO}\,,
\eeq
and can be conveniently split in the sum of the mixed spin1-spin2 term $\hat H_{\rm S_1S_2}^{\rm ADM}$ (known up the NNLO term included) 
\beq
\hat H_{\rm S_1S_2}^{\rm ADM}=\hat H_{\rm S_1S_2}^{\rm ADM,LO}+\hat H_{\rm S_1S_2}^{\rm ADM,NLO}+\hat H_{\rm S_1S_2}^{\rm ADM,NNLO}\,,
\eeq
and the spin-squared term $\hat H_{\rm S_{1,2}^2 }^{\rm ADM}$ (known up the NLO term included)
\beq
\hat H_{\rm S_{1,2}^2}^{\rm ADM}=\hat H_{\rm S_{1,2}^2}^{\rm ADM,LO}+\hat H_{\rm S_{1,2}^2}^{\rm ADM,NLO}\,.
\eeq
Actually one has also in this case a NNLO knowledge, but in the Effective-Field-Theory (EFT) picture, which to the best of our knowledge has not been translated in ADM yet \cite{Levi:2016ofk}.

We list below for completeness all these contributions by using the associated dimensionless variables (with $c=1$)
\begin{eqnarray}
\label{replacements}
&&r=\frac{R}{GM}\,,\quad L=\frac{P_\phi}{G M\mu}\,,\quad
p_r=\frac{P_r}{\mu}\,,\nonumber\\
&& 
S_1\to \frac{S_1}{G M\mu}\,,\quad S_2\to \frac{S_2}{ G M\mu}\,, 
\end{eqnarray}
as well as the notation
\beq
p^2=\frac{L^2}{r^2}+p_r^2\,.
\eeq
The orbital and the spin-orbit parts are given by
\begin{widetext}
\begin{eqnarray}
\hat H_{\rm orb}^{\rm  N }&=& \frac12 p^2  -\frac{1}{r}
\,, \nonumber\\
\hat H_{\rm orb}^{\rm ADM, 1PN }&=&  \frac{1}{8}(3\nu-1) p^4-\frac12 \frac{L^2}{r^3}(3+\nu)+\frac{1}{2 r^2} -\frac12 \frac{p_r^2}{r}(3+2\nu)
\,, \nonumber\\
\hat H_{\rm orb}^{\rm ADM, 2PN }&=&  \frac{1}{16} p^6-\frac{5}{16}\nu p^6+\frac{5}{16}\nu^2p^6+\frac{1}{r}\left[-\frac38 p_r^4\nu^2-\frac14 p_r^2 p^2\nu^2+\left(\frac{5}{8} -\frac{5}{2} \nu-\frac{3}{8}\nu^2\right)p^4\right]\nonumber\\
&&
+\frac{1}{r^2}\left[\frac{3}{2}p_r^2\nu+\left(\frac52+4\nu  \right) p^2\right]
-\frac{1}{r^3}\left(\frac14+\frac34\nu \right)
\,,\nonumber\\
\hat H_{\rm orb}^{\rm ADM, 3PN }&=& p^8 \left(-\frac{5}{128}+\frac{35}{128}\nu-\frac{35}{64}\nu^2+\frac{35}{128}\nu^3\right)\nonumber\\
&&+\frac{1}{r} \left[-\frac{5}{16}\nu^3 p_r^6
+\left(\frac{3}{16} \nu^2
-\frac{3}{16} \nu^3\right)p_r^4 p^2
+\left(\frac{1}{8} \nu^2-\frac{3}{ 16} \nu^3\right)p_r^2p^4
+\left(-\frac{7}{16} +\frac{21}{8}\nu -\frac{53}{16}\nu^2 -\frac{5}{16}\nu^3 \right)p^6 \right]\nonumber\\
&&+\frac{1}{r^2}\left[ \left(\frac{5}{12}\nu +\frac{43}{12}\nu^2\right) p_r^4
+\left(\frac{17}{16}\nu+\frac{15}{8} \nu^2\right)p_r^2  p^2+\left(-\frac{27}{16} +\frac{17}{2}  \nu+\frac{109}{16}\nu^2  \right)p^4\right]\nonumber\\
&& +\frac{1}{r^3}\left[\left(-\frac{85}{16}\nu-\frac{7}{4} \nu^2-\frac{3}{64}\nu\pi^2\right)p_r^2+\left(-\frac{25}{8}   -\frac{335}{48}\nu  -\frac{23}{8}\nu^2  +\frac{1}{64}\nu \pi^2 \right)p^2\right]\nonumber\\
&&+\frac{1}{r^4}\left(\frac{1}{8}+\frac{109}{12}\nu-\frac{21}{32}\nu\pi^2\right)\,,
\end{eqnarray}
and
\begin{eqnarray}
\hat H_{\rm SO}^{\rm ADM,LO}&=& \left(\frac12 \nu+\frac{3}{4}+\frac{3}{4}\Delta\right) \frac{L}{r^3} S_1
+1\leftrightarrow 2
\,,\nonumber\\
\hat H_{\rm SO}^{\rm ADM,NLO}&=&  \left\{\left[ \left(-\frac{5}{16}+\nu\right)\Delta-\frac{5}{16}+\frac{13}{8} \nu+\frac38  \nu^2\right] \frac{L^2}{r^2}
+\left[ \left(-\frac{5}{16}+\frac{11}{8}\nu\right)\Delta-\frac{5}{16}+2 \nu+\frac{9}{8} \nu^2\right] p_r^2   \right.\nonumber\\
&&\left.
+\left[\left(-\frac52-\nu\right)\Delta-2 \nu-\frac52\right]\frac{1}{r}  \right\} \frac{L}{r^3} S_1
+1\leftrightarrow 2
\,,\nonumber\\
\hat H_{\rm SO}^{\rm ADM,NNLO}&=&
\left\{\left[
\left(-\frac{37}{32} \nu+\frac{7}{32}+\frac{39}{32}\nu^2\right)\Delta-\frac{51}{32}\nu+\frac{5}{16}\nu^3+\frac{77}{32}\nu^2+\frac{7}{32}\right]\frac{L^4}{r^4}\right. \nonumber\\
&&+\left[\left(\frac{7}{16}-\frac{83}{32}\nu+3\nu^2\right)\Delta-\frac{111}{32}\nu+\frac{7}{16}+\frac{19}{16}\nu^3+\frac{83}{16}\nu^2\right]\frac{L^2p_r^2}{r^2}
\nonumber\\
&&  +
\left[\left(-\frac{129}{16} \nu-\frac{39}{32}\nu^2+\frac{27}{16}\right)\Delta+\frac{27}{16}-\frac{183}{16}\nu-\frac{151}{32}\nu^2 \right]\frac{L^2}{r^3}  \nonumber\\
&&+\left[\left(\frac{7}{32}+\frac{57}{32} \nu^2-\frac{23}{16}\nu\right)\Delta+\frac{7}{32}+\frac{29}{16}\nu^3-\frac{15}{8}\nu+\frac{89}{32}\nu^2\right] p_r^4 
\nonumber\\
&& 
+\left[\left(\frac{27}{16}-\frac{177}{16}\nu-\frac{21}{8}\nu^2\right)\Delta-\frac{231}{16}\nu-\frac{129}{8}\nu^2+\frac{27}{16}\right]\frac{p_r^2}{r} \nonumber\\
&&\left. +\left[\left(\frac{75}{16}+\frac{41}{8} \nu\right)\Delta+\frac{75}{16}+\frac14 \nu^2+\frac{25}{4}\nu \right]\frac{1}{r^2}
 \right\} \frac{L}{r^3}S_1
+1\leftrightarrow 2
\,,
\end{eqnarray}
respectively, where the symbol $1\leftrightarrow 2$ stands for all the spin-dependent terms with the particle labels 1 and 2 exchanged ($S_1\leftrightarrow S_2$ and  $\Delta\leftrightarrow-\Delta$).

Finally, the spin1-spin2 part and the spin-squared part are given by
\begin{eqnarray}
\hat H_{\rm S_1S_2}^{\rm ADM,LO}&=& -\frac{\nu}{r^3} S_1 S_2
\,,\nonumber\\
\hat H_{\rm S_1S_2}^{\rm ADM,NLO}&=&\left[  \left(-\frac32  \nu-\nu^2\right)\frac{L^2}{r^2} 
+ \left(\frac32 \nu-\frac74\nu^2\right) p_r^2  +\frac{6}{r}\nu  \right] \frac{1}{r^3}S_2 S_1
\,, \nonumber\\
\hat H_{\rm S_1S_2}^{\rm ADM,NNLO}&=& \left\{
\left(-\frac{23}{8} \nu^2-\frac{7}{8} \nu^3+\frac{9}{8} \nu\right)\frac{ L^4}{r^4}  
+\left[\left(\frac{19}{16} \nu^2-\frac{13}{4} \nu^3\right) p_r^2  
+\left(\frac{47}{2} \nu^2+9 \nu\right)\frac{1}{r}\right]  \frac{L^2}{r^2} \right.\nonumber\\
&&+\left.
\left[\left(-\frac{19}{8} \nu^3-\frac{9}{8} \nu+\frac{65}{16} \nu^2\right) p_r^4  +\left(-9 \nu+\frac{69}{4} \nu^2\right)\frac{p_r^2}{r}+\left(-\frac{63}{4} \nu-\frac{19}{4} \nu^2\right)\frac{1}{r^2} \right]\right\} \frac{1}{r^3} S_2 S_1 
\,,
\end{eqnarray}
and
\begin{eqnarray}
\hat H_{\rm S_{1,2}^2}^{\rm ADM,LO}&=&  \left(\frac{1}{2}\nu-\frac{1}{4}-\frac{1}{4} \Delta\right) \frac{1}{r^3}S_1^2
+1\leftrightarrow 2
\,,\nonumber\\
\hat H_{\rm S_{1,2}^2}^{\rm ADM,NLO}&=& \left[\left(-\frac32 \nu+\frac98 \nu^2+\frac{3}{16}-\frac{9}{8}\Delta \nu+\frac{3}{16}\Delta\right) p_r^2  \right. \nonumber\\
&&\left. +\left(\frac32 \Delta-\frac{13}{4} \nu+\frac32 -\frac{1}{4}\Delta \nu  \right)\frac{1}{r} 
+\left(-\frac38 \Delta \nu-\frac38   \nu\right)\frac{L^2}{r^2} \right]\frac{1}{r^3} S_1^2
+1\leftrightarrow 2
 \,,
\end{eqnarray}
\end{widetext}
respectively.

\subsection{Computing the redshift invariant}

The redshift invariant is defined as
\beq
\label{z1_def}
  z_1  =\frac{\partial H}{\partial m_1}\,,
\eeq
where all phase-space variables (except to $m_1$) are kept as constant, so that one needs the total ADM Hamiltonian \eqref{Hadmdef} (i.e., including the rest energy of the two bodies), with the physical units fully restored according to the relations \eqref{replacements}. 
Following Ref. \cite{Barack:2011ed} one then computes its orbital average 
\beq
\label{z1_average}
\langle z_1 \rangle_t =\frac{1}{T_r} \oint z_1 dt\,,
\eeq
over a radial period, which is a gauge-invariant quantity. 
It is useful to introduce the new ADM radial variable parametrization along eccentric (equatorial) orbits
\beq
r=\frac{1}{u (1+e\cos\chi)} \,,
\eeq
where $u$ denotes the reciprocal of the semi-latus rectum and $e$ the eccentricity. Both such quantities are coordinate-dependent and hence gauge-dependent.
In order to compare the results with those of the previous section one has to express the redshift function in terms of gauge-invariant variables.
We will proceed as follows.
All quantities used in the calculation are expanded both in PN sense, i.e., in powers of $\eta=1/c$, and in the spin variables up to the second order.

First of all, from the energy conservation $\hat H^{\rm ADM}=E$ one obtains $p_r$ as a function of $E$, $L$ and $u$.
Bound orbits at the periastron ($\chi=0$) and apoastron ($\chi=\pi$) are characterized by the vanishing of the radial component of the spatial momentum, i.e., $p_r=0$, leading to the relations $E=E(u,e)$ and $L=L(u,e)$.
The latter can then be inverted as
\beq
\label{uevsEL}
u=u(E,L)\,,\qquad e=e(E,L)\,,
\eeq
which allow one to express the gauge-dependent quantities $u$ and $e$ in terms of the gauge-invariant (physical) variables $E$ and $L$.

Next one determines the fundamental frequencies of the motion and associated periods, namely those of the radial and azimuthal motions
\begin{eqnarray}
T_r&=& \oint dt 
=\oint \left(\frac{\partial H}{\partial p_r}\right)^{-1}dr =2\int_0^\pi  \left(\frac{\partial H}{\partial p_r}\right)^{-1}\frac{dr}{d\chi}d\chi
\,, \nonumber\\
\Phi&=&\oint d\phi=\oint \frac{\partial H}{\partial L}dt = 2\int_0^\pi  \frac{\partial H}{\partial L}\left(\frac{\partial H}{\partial p_r}\right)^{-1} \frac{dr}{d\chi}d\chi\,,\nonumber\\
\end{eqnarray}
with
\beq
\Omega_r=\frac{2\pi}{T_r}\,, \qquad
\Omega_\phi=\frac{\Phi}{T_r}\,,
\eeq
and finally the averaged value \eqref{z1_average} of the redshift as a function of the gauge-dependent variables $u$ and $e$.  
The latter should then be re-expressed in terms of a pair of gauge invariant variables, e.g., the total energy and angular momentum though Eq. \eqref{uevsEL}.
A convenient choice is 
\beq
\label{newvars}
\hat k=\frac{k}3\,,\qquad \iota =\frac{x}{\hat k}\,,
\eeq
which are simply related to the (fractional) periastron advance per radial period $k=\frac{\Phi}{2\pi}-1$ and the dimensionless azimuthal frequency $x = (M\Omega_\phi)^{2/3}$.
Computing these two quantities allows one to express $u$ and $e$ in terms of $\hat k$ and $\iota$, or equivalently $\iota$ and $x$. 
The transformation reads
\begin{eqnarray}
u(\iota,x)&=&u_{\rm orb}(\iota,x)+u_{\rm SO}(\iota,x)+u_{\rm SS}(\iota,x)
\,,\nonumber\\
e^2(\iota,x)&=&e^2_{\rm orb}(\iota,x)+e^2_{\rm SO}(\iota,x)+e^2_{\rm SS}(\iota,x)
\,,
\end{eqnarray}
with
\begin{eqnarray}
u_{\rm SS}(\iota,x)&=&u_{\rm S_1S_2}(\iota,x)+u_{\rm S_{1,2}^2}(\iota,x)
\,,\nonumber\\
e^2_{\rm SS}(\iota,x)&=&e^2_{\rm S_1S_2}(\iota,x)+e^2_{\rm S_{1,2}^2}(\iota,x)
\,,
\end{eqnarray}
whose first terms are listed below
\begin{widetext}
\begin{eqnarray}
u_{\rm orb}(\iota,x)&=&\frac{x}{\iota}
+\left[\left(-\frac34+\frac12\nu\right)\frac{1}{\iota}+\left(-\frac{11}{4}+\frac32\nu\right)\frac{1}{\iota^2}\right] x^2
\nonumber\\
&&
+\left[\left(\frac{5}{24}\nu^2+\frac{9}{8}\nu-\frac{27}{16}\right)\frac{1}{\iota}
+\left(-\frac{61}{3}\nu+\frac32 \nu^2+\frac{41}{128}\nu\pi^2+\frac{59}{16}\right)\frac{1}{\iota^2}\right.\nonumber\\
&&\left.
+\left(\frac{247}{6}\nu+\frac{39}{8}+\frac{33}{8}\nu^2-\frac{205}{128}\nu\pi^2\right)\frac{1}{\iota^3}\right]x^3
+O(x^4)\,,
\nonumber\\
u_{\rm SO}(\iota,x)&=&
\left\{
\left[-\frac23 \Delta +\frac23-\frac13\nu\right] \frac{x^{3/2}}{\iota^{3/2}}
+\left\{
\left[\left(-\frac12\nu+\frac14\right)\Delta-\frac14\nu^2-\frac14+\frac{13}{8}\nu
\right]\frac{1}{\iota^{3/2}}\right.\right.\nonumber\\
&&\left.\left.
 +\left[\left(-\frac{11}{4}\nu+\frac{29}{12}\right)\Delta-\frac{29}{12}+\frac{11}{24}\nu-\frac{7}{4}\nu^2
\right]\frac{1}{\iota^{5/2}}\right\}x^{5/2}
+O(x^{7/2})\right\}\chi_1
+1\leftrightarrow2\,,
\nonumber\\
u_{\rm S_1S_2}(\iota,x)&=&
\left\{
\left(3\nu+\frac13\nu^2\right)\frac{x^2}{\iota^2}
+\left[\left(\frac13\nu^3+\frac32\nu^2+\frac{11}{4}\nu\right)\frac{1}{\iota^2}
+\left(\frac{469}{18}\nu^2+\frac{10}{3}\nu^3-\frac{59}{2}\nu\right)\frac{1}{\iota^3}\right]x^3
+O(x^4)\right\}\chi_1\chi_2
\,,
\nonumber\\
u_{\rm S_{1,2}^2}(\iota,x)&=&\left\{
\left[\left(-\frac{13}{12}+\frac23 \nu\right)\Delta +\frac16 \nu^2-\frac{17}{6}\nu+\frac{13}{12}\right]\frac{x^2}{\iota^2}
\right.
\nonumber\\
&&
+\left\{
\left[\left(\frac23\nu^2-\frac{53}{24}\nu-\frac{1}{16}\right)\Delta 
+\frac{25}{12}\nu+\frac16 \nu^3+\frac{1}{16}-\frac{43}{12}\nu^2\right]\frac{1}{\iota^2}\right.\nonumber\\
&&
\left.\left.
+\left[\left(\frac{17}{3}\nu^2+\frac{211}{36}-\frac{50}{9}\nu\right)\Delta
-\frac{211}{36}-\frac{677}{36}\nu^2+\frac{5}{3}\nu^3+\frac{311}{18}\nu\right]\frac{1}{\iota^3}\right\}x^3+O(x^4)\right\}\chi_1^2+1\rightarrow 2\,,
\end{eqnarray}
\begin{eqnarray}
e^2_{\rm orb}(\iota,x)&=&1-\iota
+\left[-\frac34 +\frac32\nu+\left(\frac56\nu-\frac{15}{4}\right)\iota\right] x
\nonumber\\
&&
+\left[\left(\frac{15}{8}\nu^2+\frac{557}{12}\nu-\frac{205}{128}\nu\pi^2-\frac{35}{16}\right)\frac{1}{\iota}
+\left(-\frac{5}{24}\nu^2-\frac{25}{2}+\frac{20}{3}\nu\right)\iota 
+(2\nu-5)\iota^{1/2}\right.\nonumber\\
&&\left.
+\frac{9}{16}-\frac{19}{2}\nu-\frac12 \nu^2+\frac{41}{128}\nu\pi^2\right] x^2
+O(x^3)\,,
\nonumber\\
e^2_{\rm SO}(\iota,x)&=&\left\{
\left(-\frac23\Delta+\frac23-\frac13\nu\right)\iota^{1/2} x^{1/2}\right.
\nonumber\\
&&\left.
+\left\{\left[\left(-\frac{15}{4}+\frac{7}{18}\nu\right)\Delta 
+\frac{7}{36}\nu^2-\frac{91}{72}\nu+\frac{15}{4}\right]\iota^{1/2}
+\left[\left(\frac{1}{12}-\frac34 \nu\right)\Delta-\frac{1}{12}-\frac{65}{24}\nu-\frac34\nu^2
\right]\frac{1}{\iota^{1/2}}\right\} x^{3/2}
\right\}\chi_1\nonumber\\
&&
+O(x^{5/2})+1\leftrightarrow 2\,,
\nonumber\\
e^2_{\rm S_1S_2}(\iota,x)&=&\left\{
\left(\frac19 \nu^2+\frac13 \nu\right)x
+\left[-\frac{1}{27}\nu^3+\frac{1}{18}\nu^2-\frac{17}{12}\nu
+\left(\nu^3+\frac{43}{9}\nu^2-7\nu\right)\frac{1}{\iota}\right]x^2
+O(x^3)
\right\}\chi_1\chi_2
\nonumber\\
e^2_{\rm S_{1,2}^2}(\iota,x)&=&\left\{
\left[\left(\frac29 \nu-\frac{7}{36}\right)\Delta
+\frac{7}{36}-\frac{11}{18}\nu+\frac{1}{18}\nu^2\right]x\right.
\nonumber\\
&&
+\left\{
\frac{1}{48}-\frac{1}{54}\nu^3-\frac{53}{54}\nu+\frac{31}{108}\nu^2
+\left(-\frac{2}{27}\nu^2-\frac{1}{48}+\frac{203}{216}\nu\right)\Delta\right.\nonumber\\
&&\left.\left. 
+\left[\left(\frac{85}{72}-\frac{31}{36}\nu+\frac32 \nu^2\right)\Delta 
-\frac{85}{72}+\frac12 \nu^3+\frac{29}{9}\nu-\frac{28}{9}\nu^2\right]\frac{1}{\iota}
\right\}x^2
+O(x^3)
\right\}\chi_1^2
+1\leftrightarrow 2\,,
\end{eqnarray}
where we have used the spin variables $\chi_1$ and $\chi_2$ instead of $S_1$ and $S_2$. 
The redshift invariant as a function of $\iota$ and $x$ then turns out to be
\beq
\langle z_1 \rangle_t(\iota,x)=\langle z_1 \rangle_{t\,\rm orb}(\iota,x)+\langle z_1 \rangle_{t\,\rm SO}(\iota,x)+\langle z_1 \rangle_{t\,\rm SS}(\iota,x)\,,
\eeq
with
\begin{eqnarray}
\langle z_1 \rangle_{t\,\rm orb}(\iota,x)&=&1
+\left(
-\frac34 +\frac12 \nu-\frac34 \Delta 
\right)x\nonumber\\
&+&
\left[
 \left(\frac{15}{16}-\frac{1}{8}\nu\right)\Delta+\frac12\nu+\frac{15}{16}+\frac{5}{24}\nu^2+(-3-3\Delta)\frac{1}{\iota^{1/2}}
+\left(\frac32 -\nu+\frac32\Delta\right)\frac{1}{\iota}
\right]x^2\nonumber\\
&+&
\left\{
\left(\frac{5}{32}-\frac{1}{32}\nu^2-\frac{5}{16}\nu\right)\Delta+\frac{9}{32}\nu^2+\frac{1}{16}\nu^3+\frac{5}{32}
+\left[\left(-\frac32 \nu+\frac{15}{8}\right)\Delta+2\nu^2+\frac{15}{8}+\nu\right]\frac{1}{\iota^{1/2}}\right.\nonumber\\
&&
+\left[\left(-\frac{15}{4}+\frac12\nu\right)\Delta -\frac{15}{4}-2\nu-\frac{5}{6}\nu^2\right]\frac{1}{\iota}\nonumber\\
&&\left.
+\left[\left(\frac{37}{8}+\frac52\nu\right)\Delta+\frac52\nu-5\nu^2+\frac{37}{8}\right]\frac{1}{\iota^{3/2}}
+\left(\frac52\nu-\frac{15}{4}\Delta -\frac{15}{4}\right)\frac{1}{\iota^2}
\right\}x^3\nonumber\\
&+&\left\{
\left(-\frac{5}{128}\nu^2-\frac{7}{1728}\nu^3-\frac{45}{128}\nu-\frac{37}{256}\right)\Delta-\frac{37}{256}+\frac{49}{864}\nu^3-\frac56\nu+\frac{5}{64}\nu^2+\frac{91}{10368}\nu^4\right.\nonumber\\
&&
+\left[\left(\frac{315}{128}-\frac{35}{16}\nu-\frac{21}{32}\nu^2\right)\Delta+\frac76\nu^3+\frac{315}{128}+\frac{35}{8}\nu+\frac{77}{96}\nu^2\right]\frac{1}{\iota^{1/2}}\nonumber\\
&&
+\left[\left(\frac{3}{16}\nu^2-\frac{15}{16}+\frac{15}{8}\nu\right)\Delta-\frac38\nu^3-\frac{15}{16}-\frac{27}{16}\nu^2\right]\frac{1}{\iota}\nonumber\\
&&
+\left[\left(\frac{41}{128}\nu\pi^2-\frac{285}{64}-\frac{211}{12}\nu+\frac{45}{16}\nu^2\right)\Delta\right.\nonumber\\
&&\left.
+\frac{3781}{144}\nu^2-\frac{41}{96}\nu^2\pi^2-\frac{1129}{48}\nu-\frac{285}{64}+\frac{41}{128}\nu\pi^2-\frac{25}{4}\nu^3\right]\frac{1}{\iota^{3/2}}\nonumber\\
&&
+\left[\left(-\frac{7}{4}\nu+\frac{105}{8}\right)\Delta+\frac{35}{12}\nu^2+\frac{105}{8}+7\nu\right]\frac{1}{\iota^2}\nonumber\\
&&
+\left[\left(-\frac{123}{128}\nu\pi^2-\frac{1797}{128}+\frac{99}{32}\nu^2+\frac{355}{16}\nu\right)\Delta\right.\nonumber\\
&&\left.
-\frac{1797}{128}+\frac{123}{64}\nu^2\pi^2-\frac{1321}{32}\nu^2-\frac{123}{128}\nu\pi^2-\frac{33}{4}\nu^3+\frac{355}{16}\nu\right]\frac{1}{\iota^{5/2}}\nonumber\\
&&\left.
+\left(10-\frac{20}{3}\nu+10\Delta\right)\frac{1}{\iota^3}
\right\}x^4
+O(x^5)
\,,
\end{eqnarray} 
\begin{eqnarray}
\langle z_1 \rangle_{t\,\rm SO}(\iota,x)&=&
\left[
 \left(\chi_1\nu+\left(1-\frac32\nu\right)\chi_2\right)\Delta +\left(-\frac72\nu+1+\nu^2\right)\chi_2+(\nu^2-\nu)\chi_1 
\right]\frac{x^{5/2}}{\iota}\nonumber\\
&+&
\left\{
\left[\left[\left(\frac{25}{24}\nu^2-\frac{7}{12}\nu\right)\chi_1+\left(\frac{95}{24}\nu-\frac{37}{24}\nu^2-\frac{7}{4}\right)\chi_2\right]\Delta\right.\right.\nonumber\\
&&\left.
+\left(\frac{7}{12}\nu+\frac{5}{6}\nu^3-\frac{83}{24}\nu^2\right)\chi_1
+\left(-\frac{7}{4}-\frac{161}{24}\nu^2+\frac{5}{6}\nu^3+\frac{179}{24}\nu\right)\chi_2\right]\frac{1}{\iota}\nonumber\\
&&
+\left[\left[\left(\frac{25}{8}\nu^2-\frac54\nu\right)\chi_1+\left(-\frac54-\frac{39}{8}\nu^2+\frac{13}{8}\nu\right)\chi_2\right]\Delta\right.\nonumber\\
&&\left.\left.
+\left(\frac72\nu^3+\frac{19}{8}\nu^2+\frac54 \nu\right)\chi_1+\left(-\frac{45}{8}\nu^2+\frac72\nu^3-\frac54+\frac{33}{8}\nu\right)\chi_2\right]\frac{1}{\iota^2}
\right\}x^{7/2}\nonumber
\end{eqnarray}
\begin{eqnarray}
&+&
\left\{
\left[\left[\left(\frac{15}{16}\nu+\frac{95}{144}\nu^3+\frac{17}{36}\nu^2\right)\chi_1
+\left(\frac{77}{72}\nu^2-\frac{45}{16}+\frac{253}{96}\nu-\frac{35}{36}\nu^3\right)\chi_2\right]\Delta
\right.\right.\nonumber\\
&&\left.
+\left(-\frac{181}{144}\nu^3+\frac{35}{72}\nu^4-\frac{15}{16}\nu-\frac{275}{144}\nu^2\right)\chi_1
+\left(-\frac{45}{16}-\frac{31}{9}\nu^3+\frac{35}{72}\nu^4-\frac{245}{144}\nu^2+\frac{793}{96}\nu\right)\chi_2\right]\frac{1}{\iota}\nonumber\\
&&
+\left[
\left[\left(-\frac{2701}{144}\nu^2+\frac{205}{48}\nu^3+\frac{1025}{1536}\nu^2\pi^2+\frac{35}{48}\nu\right)\chi_1\right.\right.\nonumber\\
&&\left.
+\left(-\frac{1517}{1536}\nu^2\pi^2+\frac{1207}{36}\nu^2-\frac{331}{48}\nu^3+\frac{35}{16}-\frac{2039}{96}\nu+\frac{41}{64}\nu\pi^2\right)\chi_2\right]\Delta\nonumber\\
&&
+\left(-\frac{35}{48}\nu-\frac{287}{1536}\nu^2\pi^2+\frac{14}{3}\nu^4-\frac{2873}{144}\nu^3+\frac{1237}{144}\nu^2+\frac{205}{384}\nu^3\pi^2\right)\chi_1\nonumber\\
&&\left.
+\left(\frac{205}{384}\nu^3\pi^2+\frac{2635}{36}\nu^2+\frac{41}{64}\nu\pi^2-\frac{4709}{144}\nu^3-\frac{3485}{1536}\nu^2\pi^2+\frac{35}{16}-\frac{2459}{96}\nu+\frac{14}{3}\nu^4\right)\chi_2\right]\frac{1}{\iota^2}\nonumber\\
&&
+\left[
\left[\left(\frac{5}{8}\nu+\frac{2135}{24}\nu^2+\frac{105}{8}\nu^3-\frac{2255}{512}\nu^2\pi^2\right)\chi_1\right.\right.\nonumber\\
&&\left.
+\left(\frac{5}{8}-\frac{257}{2}\nu^2+\frac{949}{12}\nu-\frac{165}{8}\nu^3-\frac{205}{64}\nu\pi^2+\frac{3075}{512}\nu^2\pi^2\right)\chi_2\right]\Delta\nonumber\\
&&
+\left(\frac{2213}{24}\nu^3+15\nu^4-\frac{1691}{24}\nu^2+\frac{1025}{512}\nu^2\pi^2-\frac58 \nu-\frac{205}{64}\nu^3\pi^2\right)\chi_1\nonumber\\
&&\left.\left.
+\left(15\nu^4+\frac{6355}{512}\nu^2\pi^2-\frac{205}{64}\nu^3\pi^2+\frac58 -\frac{3455}{12}\nu^2+\frac{467}{6}\nu-\frac{205}{64}\nu\pi^2+\frac{1403}{24}\nu^3\right)\chi_2\right]\frac{1}{\iota^3}
\right\}x^{9/2}\nonumber\\
&&
+O(x^{11/2})
\,,
\end{eqnarray} 
and
\begin{eqnarray}
\langle z_1 \rangle_{t\,\rm SS}(\iota,x)&=&
\left\{
\left[\left(\frac{5}{4}\nu-\frac{25}{24}\nu^2\right)\chi_1^2+\frac{1}{12}\nu^2\chi_1\chi_2+\left(\frac{5}{12}+\frac{9}{8}\nu^2-\frac{23}{12}\nu\right)\chi_2^2\right]\Delta\right.\nonumber\\
&&\left.
+\left[
\left(-\frac54 \nu-\frac13\nu^3+\frac{79}{24}\nu^2\right)\chi_1^2
+\left(-\frac{2}{3}\nu^3-\frac{23}{12}\nu^2\right)\chi_2\chi_1
+\left(\frac{33}{8}\nu^2+\frac{5}{12}-\frac{11}{4}\nu-\frac13\nu^3\right)\chi_2^2\right] 
\right\}\frac{x^3}{\iota^{3/2}}\nonumber\\
&+&
\left\{
 \left[\left[\left(-\frac{13}{12}\nu^3+\frac{245}{288}\nu+\frac{781}{576}\nu^2\right)\chi_1^2
+\left(\frac{1}{8}\nu^3+\frac{13}{96}\nu^2\right)\chi_2\chi_1
+\left(\frac{29}{24}\nu^3-\frac{17}{288}\nu+\frac{35}{96}-\frac{1711}{576}\nu^2\right)\chi_2^2\right]\Delta\right.\right.\nonumber\\
&&
+\left(-\frac{179}{576}\nu^2+\frac{571}{144}\nu^3-\frac13\nu^4-\frac{245}{288}\nu\right)\chi_1^2
+\left(-\frac{7}{9}\nu^3-\frac{227}{96}\nu^2-\frac{2}{3}\nu^4\right)\chi_2\chi_1\nonumber\\
&&\left.
+\left(\frac{35}{96}-\frac{227}{288}\nu-\frac{2063}{576}\nu^2-\frac13\nu^4+\frac{805}{144}\nu^3\right)\chi_2^2\right]\frac{1}{\iota^{3/2}}\nonumber\\
&&
+\left[\left[\left(-\frac{185}{24}\nu^3+\frac{295}{64}\nu^2-\frac{85}{32}\nu\right)\chi_1^2
+\left(\frac{283}{96}\nu^2+\frac{23}{24}\nu^3\right)\chi_2\chi_1\right.\right.\nonumber\\
&&\left.
+\left(-\frac{85}{96}-\frac{757}{64}\nu^2+\frac{26}{3}\nu^3+\frac{467}{96}\nu\right)\chi_2^2\right]\Delta\nonumber\\
&&
+\left(\frac{85}{32}\nu-\frac{23}{8}\nu^4-\frac{497}{64}\nu^2+\frac{217}{12}\nu^3\right)\chi_1^2
+\left(-\frac{181}{8}\nu^3+\frac{1363}{96}\nu^2-\frac{23}{4}\nu^4\right)\chi_2\chi_1
\nonumber\\
&&\left.\left.
+\left(-\frac{3799}{192}\nu^2-\frac{85}{96}-\frac{23}{8}\nu^4+\frac{637}{96}\nu+\frac{679}{24}\nu^3\right)\chi_2^2
\right]\frac{1}{\iota^{5/2}} 
\right\}x^4
+O(x^5)
\,.
\end{eqnarray} 
\end{widetext}
The orbital part has been computed in Ref. \cite{Akcay:2015pza} (see Eqs. (4.42a)--(4.42d)).
Notice that the 3PN contribution ${\mathcal V}_{\rm 3PN}$ is misprinted there. In fact, there are two missing terms proportional to $\iota^{-2}$ and $\iota^{-3}$, necessary to reproduce the corresponding terms proportional to $\lambda^{-2}$ and $\lambda^{-3}$ in the correct 1SF expansion (4.50b) there.
${\mathcal V}_{\rm 3PN}$ then should read
\begin{eqnarray}
{\mathcal V}_{\rm 3PN}&=&{\mathcal V}_{\rm 3PN}\vert_{{\rm Ref.} {\mbox {\scriptsize \cite{Akcay:2015pza}}}}\nonumber\\
&&
+\left[\frac{21}{8}-49\nu+\frac{7}{12}\nu^2+\left(-\frac{35}{4}\nu+\frac{21}{8}\right)\Delta\right]\frac1{\iota^2}\nonumber\\
&&
+\left[-10 +\frac{20}{3}\nu-10\Delta\right]\frac1{\iota^3}\,.
\end{eqnarray} 
Eq. (4.4) in Ref. \cite{Tiec:2015cxa} is likely to propagate this omission too.

The GSF contribution can be extracted by substituting the new variables $y=(m_2\Omega_\phi)^{2/3}$ and $\lambda=y/\hat k$, which are related to $x$ and $\iota$ by $x=y(1+q)^{2/3}$ and $\iota=\lambda(1+q)^{2/3}$, respectively, into the previous expressions, expanding them in power series of the mass ratio $q$ and selecting the first order terms.
One then gets the 1SF part 
\begin{widetext}
\begin{eqnarray}
\langle z_1 \rangle_{t\,\rm orb}^{\rm 1SF}(y,\lambda)&=& 
y+\left(-\frac{2}{\lambda}+1\right)y^2
+\left(\frac{2}{\lambda^{1/2}}+\frac{5}{\lambda^2}+\frac{5}{\lambda^{3/2}}-\frac{4}{\lambda}\right)y^3\nonumber\\
&&
+\left[-\frac{5}{3}+\frac{35}{4\lambda^{1/2}}+\left(-\frac{1129}{24}+\frac{41}{64}\pi^2\right)\frac1{\lambda^{3/2}}+\frac{14}{\lambda^2}+\left(-\frac{123}{64}\pi^2+\frac{355}{8}\right)\frac1{\lambda^{5/2}}-\frac{40}{3\lambda^3}\right]y^4
+O(y^5)
\,,\nonumber\\
\langle z_1 \rangle_{t\,\rm S}^{\rm 1SF}(y,\lambda)&=& 
\left\{-\frac{5}{\lambda}y^{5/2}+\left(\frac{109}{12\lambda}+\frac{23}{4\lambda^2}\right)y^{7/2}
+\left[\frac{163}{48\lambda}+\left(-\frac{703}{16}+\frac{41}{32}\pi^2\right)\frac1{\lambda^2}+\left(\frac{1883}{12}-\frac{205}{32}\pi^2\right)\frac1{\lambda^3}\right]y^{9/2}\right\}\chi_2\nonumber\\
&&
+O(y^{11/2})
\,,\nonumber\\
\langle z_1 \rangle_{t\,\rm SS}^{\rm 1SF}(y,\lambda)&=& 
\left[-\frac{14}{3\lambda^{3/2}}y^3+\left(\frac{23}{2\lambda^{5/2}}-\frac{13}{36\lambda^{3/2}}\right)y^4\right]\chi_2^2
+O(y^5)\,.
\end{eqnarray}
The last step consists in computing the Kerr background values for $y$ and $\lambda$, both functions of $u_p$ and $e_p$ (say, to distinguish them from the corresponding ADM quantities $u$ and $e$), and substituting them into the previous 1SF expressions. 
Setting $\chi_2=\hat a$ finally gives
\begin{eqnarray}
\langle z_1 \rangle_{t\,\rm orb}^{\rm 1SF}(u_p,e_p)&=&  (1-e_p^2) u_p+(-1+2e_p^2-e_p^4) u_p^2+\left(-1+5 e_p^2-\frac{19}{4} e_p^4\right) u_p^3\nonumber\\
&+& \left[\frac{76}{3}-\frac{41}{32}\pi^2+\left(\frac{23}{3}+\frac{41}{32}\pi^2\right) e_p^2+\left(-\frac{339}{8}+\frac{123}{256}\pi^2\right) e_p^4\right] u_p^4
+O(u_p^5)
\,,\nonumber\\
\langle z_1 \rangle_{t\,\rm S}^{\rm 1SF}(u_p,e_p)&=&\left[\left(-3+\frac72 e_p^2+\frac18  e_p^4\right) u_p^{5/2}+\left(-3-\frac{55}{2} e_p^2+\frac{351}{8} e_p^4\right) u_p^{7/2}\right.\nonumber\\
&&\left.
+\left(-21-\frac{311}{2} e_p^2+\frac{2377}{16} e_p^4\right) u_p^{9/2}
+O(u_p^{11/2})\right] \hat a
\,,\nonumber\\
\langle z_1 \rangle_{t\,\rm SS}^{\rm 1SF}(u_p,e_p)&=& \left[\left(1-e_p^2-\frac14 e_p^4\right)u_p^3+\left(10+\frac{45}{2} e_p^2-\frac{393}{8} e_p^4\right) u_p^4
+O(u_p^5)\right] \hat a^2
\,,
\end{eqnarray}
\end{widetext}
which coincide with the GSF results for $\delta U=-\langle z_1 \rangle_t/z_0^2$, with $z_0=U_0^{-1}$, of the previous section.

\subsection{Circular limit}

Finally, let us discuss the circular orbit limit of previous results.
The variables $\iota$ and $x$ are not independent in this limit.
Recalling the definition \eqref{newvars}, in order to express $\iota$ as a function of $x$ it is enough to use the relation $k_{\rm circ}(x)$ for the fractional periastron advance (see Eqs. (9a)--(9h) in Ref. \cite{Tiec:2013twa})
\beq
k^{\rm circ}(x)=k^{\rm circ}_{\rm orb}(x)+k^{\rm circ}_{\rm S}(x)+k^{\rm circ}_{\rm SS}(x)\,,
\eeq
with
\beq
k^{\rm circ}_{\rm SS}(x)=k^{\rm circ}_{\rm S_1S_2}(x)+k^{\rm circ}_{S_{1,2}^2}(x)\,,
\eeq
where
\begin{widetext}
\begin{eqnarray}
k^{\rm circ}_{\rm orb}(x)&=&
3 x+\left(\frac{27}{2}-7\nu\right) x^2+\left(7\nu^2-\frac{649}{4}\nu+\frac{135}{2}+\frac{123}{32}\nu\pi^2\right) x^3
+O(x^4)
\,,\nonumber\\
k^{\rm circ}_{\rm S}(x)&=&
\left[
(-2+2\Delta+\nu)x^{3/2}
+\left(-\frac{17}{4}\Delta\nu-17-\nu^2+17\Delta +\frac{81}{4}\nu\right) x^{5/2}\right.\nonumber\\
&&
+\left. 
\left(-\frac{733}{12}\nu^2+\frac13\nu^3+\frac{11581}{48}\nu+\frac{11}{3}\Delta\nu^2-126-\frac{5317}{48}\Delta\nu+126\Delta \right)x^{7/2}
+O(x^{9/2})\right]\chi_1
+1\leftrightarrow 2 \,,
\nonumber\\
k^{\rm circ}_{\rm S_1S_2}(x)&=&
\left[
3\nu x^2+\left(2\nu^2+45\nu\right)x^3
+O(x^4)\right]\chi_2\chi_1
\,,\nonumber\\
k^{\rm circ}_{S_{1,2}^2}(x)&=&
\left[
\left(\frac34-\frac32\nu-\frac34\Delta\right)x^2
+\left(6\nu^2-\frac{189}{4}\nu+\frac{67}{4}-\frac{67}{4}\Delta+\frac{55}{4}\Delta\nu\right)x^3
+O(x^4)\right]\chi_1^2
+1\leftrightarrow 2\,,
\end{eqnarray}
so that 
\beq
\iota_{\rm circ}(x)=\frac{3x}{k_{\rm circ}(x)}\,.
\eeq
We then find 
\beq
z_1^{\rm circ}(x)=z_1{}^{\rm circ}_{\rm orb}(x)+z_1{}^{\rm circ}_{\rm S}(x)+z_1{}^{\rm circ}_{\rm SS}(x)\,,
\eeq
where
\begin{eqnarray}
z_1{}^{\rm circ}_{\rm orb}(x)&=&1
+\left(-\frac34 +\frac12 \nu-\frac34\Delta\right)x
+\left[\left(-\frac18\nu-\frac{9}{16}\right)\Delta+\frac{5}{24}\nu^2-\frac12\nu-\frac{9}{16}\right]x^2
\nonumber\\
&&
+\left[\left(-\frac{1}{32}\nu^2-\frac{27}{32}+\frac{19}{16}\nu\right)\Delta-\frac12 \nu-\frac{39}{32}\nu^2-\frac{27}{32}+\frac{1}{16}\nu^3\right] x^3
\nonumber\\
&&
+\left[\left(-\frac{41}{64}\nu\pi^2-\frac{93}{128}\nu^2+\frac{6889}{384}\nu-\frac{405}{256}-\frac{7}{1728}\nu^3\right)\Delta\right.\nonumber\\
&&\left.
+\frac{91}{10368}\nu^4+\frac{41}{192}\nu^2\pi^2-\frac{3863}{576}\nu^2+\frac{973}{864}\nu^3-\frac{41}{64}\nu\pi^2+\frac{38}{3}\nu-\frac{405}{256}\right] x^4
+O(x^5)\,,
\end{eqnarray}
\begin{eqnarray}
z_1{}^{\rm circ}_{\rm S}(x)&=&
\left[\left(\frac13\chi_1\nu+\left(-\frac56\nu+1\right)\chi_2\right)\Delta+\left(-\frac13\nu+\frac23\nu^2\right)\chi_1+\left(-\frac{17}{6}\nu+\frac23 \nu^2+1\right)\chi_2\right] x^{5/2}
\nonumber\\
&&
\left[\left(\left(-\frac{19}{18}\nu^2-\frac12\nu\right)\chi_1+\left(-\frac83 \nu+\frac{41}{36}\nu^2+\frac32\right)\chi_2\right)\Delta+\left(\frac{19}{18}\nu^2+\frac12\nu-\frac19\nu^3\right)\chi_1\right.\nonumber\\
&&\left.
+\left(\frac{179}{36}\nu^2-\frac{17}{3}\nu+\frac32-\frac19\nu^3\right)\chi_2\right] x^{7/2}
\nonumber\\
&&
+\left[ \left(\left(\frac{11}{24}\nu^3-\frac{27}{8}\nu-\frac{39}{8}\nu^2\right)\chi_1
+\left(-\frac{119}{8}\nu-\frac{19}{48}\nu^3+\frac{195}{16}\nu^2+\frac{27}{8}\right)\chi_2\right)\Delta\right.\nonumber\\
&&\left. 
+\left(\frac{27}{8}\nu-\frac{161}{24}\nu^3-\frac{39}{8}\nu^2-\frac{1}{12}\nu^4\right)\chi_1
+\left(-\frac{173}{8}\nu-\frac{1}{12}\nu^4+\frac{617}{16}\nu^2-\frac{391}{48}\nu^3+\frac{27}{8}\right)\chi_2\right] x^{9/2}\nonumber\\
&&
+O(x^{11/2})\,,
\end{eqnarray}
\begin{eqnarray}
z_1{}^{\rm circ}_{\rm SS}(x)&=&
\left[
\left(-\frac12\chi_1^2\nu+\left(\frac12\nu-\frac14\right)\chi_2^2\right)\Delta+\left(\frac12\nu-\frac12\nu^2\right)\chi_1^2+\nu^2\chi_1\chi_2+\left(-\frac14-\frac12\nu^2+\nu\right)\chi_2^2
\right]x^3
\nonumber\\
&&
\left[
\left[\left(-\frac{13}{18}\nu+\frac{137}{72}\nu^2\right)\chi_1^2-\frac{1}{12}\nu^2\chi_1\chi_2+\left(-\frac{13}{24}-\frac{179}{72}\nu^2+\frac{113}{36}\nu\right)\chi_2^2\right]\Delta\right.\nonumber\\
&&\left.
+\left(\frac{13}{18}\nu-\frac{295}{72}\nu^2+\frac{49}{36}\nu^3\right)\chi_1^2+\left(\frac{1}{12}\nu^2+\frac{7}{18}\nu^3\right)\chi_2\chi_1+\left(-\frac{553}{72}\nu^2-\frac{13}{24}+\frac{38}{9}\nu+\frac{49}{36}\nu^3\right)\chi_2^2
\right] x^4\nonumber\\
&&
+O(x^5)
\,.
\end{eqnarray}
The spin orbit terms LO and NLO in the circular case are given in Eqs. (5.5)--(5.6) of Ref. \cite{Blanchet:2012at}.
The spin square NNLO term can be obtained by using the EFT results of Ref. \cite{Levi:2014sba}, as it follows from Eq. (4.6) in Ref. \cite{Bini:2018ylh}
\begin{eqnarray}
z_1{}^{\rm circ}_{\rm SS\,,NNLO}(x)&=&
\left[
\left[\left(-\frac{67}{48}\nu+\frac{7393}{864}\nu^2-\frac{607}{108}\nu^3\right)\chi_1^2+\left(\frac{53}{72}\nu^3+\frac{143}{48}\nu^2\right)\chi_2\chi_1\right.\right.\nonumber\\
&&\left.
+\left(-\frac{67}{32}+\frac{1045}{72}\nu-\frac{10043}{432}\nu^2+\frac{1391}{216}\nu^3\right)\chi_2^2\right]\Delta\nonumber\\
&&
+\left(-\frac{649}{432}\nu^4-\frac{10291}{864}\nu^2+\frac{8453}{432}\nu^3+\frac{67}{48}\nu\right)\chi_1^2+\left(\frac{113}{48}\nu^2-\frac{583}{216}\nu^4-\frac{331}{36}\nu^3\right)\chi_2\chi_1\nonumber\\
&&\left.
+\left(-\frac{10387}{216}\nu^2-\frac{67}{32}-\frac{649}{432}\nu^4+\frac{2693}{144}\nu+\frac{7135}{216}\nu^3\right)\chi_2^2
\right] x^5
\,.
\end{eqnarray}
The corresponding 1SF expansion then reads
\begin{eqnarray}
z_1{}^{\rm circ,1SF}(y)&=& 
y-y^2-y^3+\left(\frac{76}{3}-\frac{41}{32}\pi^2\right)y^4+O(y^5)
+\left(-\frac{7}{3}y^{5/2}-\frac{13}{3}y^{7/2}-23y^{9/2}+O(y^{11/2})\right)\chi_2\nonumber\\
&&
+\left(y^3+\frac{50}{9}y^4+\frac{211}{9}y^5+O(y^6)\right)\chi_2^2
\,,
\end{eqnarray}
\end{widetext}
which agrees with the first PN terms of the corresponding 1SF expansion for $\delta U^{\rm circ}(y)=-z_1{}^{\rm circ}(y)/z_0(y)^2$ of Ref. \cite{Bini:2015xua}  for $\chi_2=\hat a$.

\section{Concluding remarks}

In a previous work we have analytically computed the GSF correction to the Detweiler-Barack-Sago redshift invariant for particles on slightly eccentric equatorial orbits around a Kerr spacetime up to the second order in both the eccentricity and spin parameter, and through the 8.5 PN order.
We have improved here its knowledge by adding terms which are fourth order in eccentricity with the same PN accuracy.
We have also checked the first terms of our final result by using the available ADM Hamiltonian for spinning binaries.
We expect that such a high-PN analytical result can be used to validate existing numerical codes on self-force calculations in a Kerr spacetime and to inform other formalisms, like the EOB model.

\end{document}